\documentclass{article}

\usepackage{amsthm, amsmath, amssymb}
\theoremstyle{plain}
\newtheorem{theorem}{Theorem}[section]
\newtheorem{lemma}[theorem]{Lemma}

\newtheorem{proposition}[theorem]{Proposition}

\newtheorem{conjecture}[theorem]{Conjecture}

\theoremstyle{definition}

\theoremstyle{remark}
\newtheorem{remark}[theorem]{Remark}

\def\squareforqed{\hbox{\rlap{$\sqcap$}$\sqcup$}}
\def\qed{\ifmmode\squareforqed\else{\unskip\nobreak\hfil
\penalty50\hskip1em\null\nobreak\hfil\squareforqed
\parfillskip=0pt\finalhyphendemerits=0\endgraf}\fi}

\newenvironment{proofof}[1]{\begin{trivlist}%
\item[]{\flushleft\em Proof of #1. }}
{\qed\end{trivlist}}

\newcommand{\thnum}[1]{{#1}^{\mathrm{th}}}
\newcommand{\defeq}{\stackrel{\mathrm{def}}{=}}
\newcommand{\entropy}{H}

\newcommand{\sepAuthor}{1in}
\newcommand{\sepAbstract}{1in}
\newcommand{\skipKeywords}{30pt}
\newcommand{\sepTitle}{2ex}

\long\def\mytitlepage#1#2#3#4{
        \thispagestyle{empty}
	\vspace*{\sepTitle}
        \begin{center}
        {\Large\bf #1}

        \vspace{\sepAuthor}
        #2\\
        \medskip

        \vspace{\sepAbstract}
        {\Large Abstract}
        \end{center}

        \noindent{#3}
        \vskip\skipKeywords

        \noindent{#4}
        \clearpage
        }
\begin{document}
\mytitlepage{
Quantum and Classical Tradeoffs\footnote{
This research was supported in part by
NSF Grant EIA-0323555.}
}
{{\large Yaoyun Shi}\\
Department of Electrical and Computer Engineering\\
The University of Michigan\\
Ann Arbor, MI 48109-2122, USA\\
E-mail: shiyy@eecs.umich.edu.}
{
We propose an approach for quantifying a quantum circuit's 
quantumness as a means to understand the nature of quantum
algorithmic speedups. Since quantum gates that do not preserve
the computational basis are necessary for achieving quantum 
speedups, it appears natural to define the quantumness of a
quantum circuit using the number of such gates.
Intuitively, a reduction in the quantumness requires
an increase in the amount of classical computation, hence giving
a ``quantum and classical tradeoff''.

In this paper we present two results on this direction. 
The first gives an asymptotic answer to the question: 
``what is the minimum number of non-basis-preserving gates required 
to generate a good approximation to a given state''.
This question is the quantum analogy of 
the following classical question, ``how many fair coins
are needed to generate a given probability distribution'',
which was studied and resolved by Knuth and Yao in 1976.
Our second result shows that any quantum algorithm 
that solves Grover's Problem of size $n$ using $k$ queries and 
$\ell$ levels of non-basis-preserving gates must have $k\cdot\ell=\Omega(n)$.
}{\noindent{\bf Keywords:} Quantum computation, quantum and classical tradeoff,
quantum state generation, quantum lower bound, Grover's Algorithm.}
\section{Introduction.}
The importance of quantum computing lies in the 
possibility that quantum mechanical algorithms
may be dramatically more efficient than the best
classical algorithms.
In order to understand the nature of 
quantum speedup, it is important to identify
features of quantum computing that are uniquely
quantum and investigate their roles in 
quantum speedups. One example of this kind of
study was taken by Jozsa and Linden \cite{JozsaL:2002:entanglement}, 
which relates the amount of entanglement 
during the computation to the difficulty of
simulating the computation. 
Our work is alone a similar line, but instead
of entanglement, we study another feature of quantum
computing: the number of gates that do not preserve
the computational basis.

It is well known that any classical
computation can be carried out, without
much sacrifice in the efficiency,
using classical reversible gates, such as
the Toffoli gate. In order to have nontrivial
quantum speedup, gates that do not preserve the
computational basis must be used.
Furthermore, the more such gates involved, 
the more difficult a straightforward classical
simulation is.

Recall that the state space of a qubit
has an orthonormal basis,
denoted by $\{|0\rangle, |1\rangle\}$,
that is fixed a priori and called
the {\em computational basis}.
The computational basis for the state
space of $n$ qubits is the tensor products of their
computational bases. Each qubit of a quantum computer
is assumed to start in the computational base state
$|0\rangle$. We follow this convention throughout this paper.

Let us formally call a gate $G$ {\em basis-changing}
if there exist two computational base vectors
$|\phi\rangle$ and  $|\psi\rangle$, such that
$|\langle \phi | G |\psi\rangle|<1$. If $G$ is not
basis-changing, $G$ is said to be {\em basis-preserving}.
An important example
of a basis-changing gate is the Hadamard gate
\[H\defeq\frac{1}{\sqrt{2}}\left(|0\rangle\langle 0|
+|1\rangle\langle 0|+
|0\rangle\langle 1|-
|1\rangle\langle 1|\right).\]
It is well known (e.g. \cite{Shi:2003:BTH}) that
any quantum circuit can be efficiently simulated
by Toffoli and Hadamard gates.
It is also easy to observe that 
a quantum circuit that uses $k$ Hadamard gates,
together with some other basis-preserving gates,
can be simulated straightforwardly by a deterministic
algorithm with a $2^k$ factor of slow-down.

Hence, it appears natural to quantify the
amount of ``quantumness'' of a quantum circuit
by the number of basis-changing gates,
and to investigate the tradeoffs between this amount of
quantumness with the best possible quantum speedup.
This is precisely the theme of our investigation.

Many interesting questions can be asked
in this diction. In particular, we present two results
in this paper.
The first is on the following question:
what is the minimum number
of basis-changing gates required to
generate a good approximation of a given quantum state?
This is in analogy to the following classical
question:
what is the least number of fair coins required
to produce a given probability distribution?
In 1976, Knuth and Yao \cite{KnuthY:1976}
solved this problem completely:
the minimum expected number of fair coins needed is
equal to the Shannon entropy of the distribution 
plus some universal constant.
We find that the answer to the quantum problem is similar.

The second result investigates the quantum-classical 
tradeoffs in solving Grover's Problem \cite{Grover:1996:Search},
also called the Unstructured Search Problem,
an important and well studied problem in quantum computing.
We prove that any quantum algorithm that solves Grovers' Problem
of size $n$
using $k$ queries and $\ell$ levels of basis-changing gates
must have $k\cdot \ell=\Omega(n)$. This tradeoff relation
is tight.

We shall present these two results in the following two sections,
followed by a discussion of open problems.
\section{Quantum State Generation.}
A classical problem studied by Knuth and Yao
in \cite{KnuthY:1976} is the following:
how many independent $0/1$ variables are needed
in order to generate a given probability distribution?
They prove that the minimum expected number of coins
is precisely $\entropy(\mathcal{D})$, the Shannon entropy of
$\mathcal{D}$, plus some universal constant. In this section 
we study the quantum analog of the question:
how many basis-changing gates are needed in order
to generate a good approximation of a
given quantum state?

For a quantum state $|\phi\rangle$, denote
by $H(\phi)$
the Shannon entropy of the probability distribution
obtained from measuring $|\phi\rangle$ in the computational
basis. We prove both upper and lower bounds to
the quantum problem in terms of $H(\phi)$.
\subsection{Upper bound.}
We first consider a special case, and then reduce the 
general case to it.
\begin{lemma}\label{lm:special}
Let $|\phi\rangle$ be a state over $n$ qubits with
nonnegative amplitudes, and $\epsilon>0$ be a real
constant.
Then there is a quantum algorithm that
uses $O(n\log \frac{n}{\epsilon})$
basis-changing gates and maps
$|0\rangle^{\otimes n}$ to a state $|\phi'\rangle$,
such that $\||\phi'\rangle - |\phi\rangle\|\le\epsilon$.
\end{lemma}
The algorithm is along the lines of the algorithm in 
\cite{Kitaev:2002:book}
for approximating an operator that maps
$|q\rangle\otimes|0\rangle^{\otimes n}$ to 
$|q\rangle\otimes\left(\sum_{j=0..q-1}\frac{1}{\sqrt{q}}|j\rangle\right)$.
\begin{proof}
Suppose 
\[|\phi\rangle=\sum_{y\in\{0, 1\}^n} \sqrt{p_y} |y\rangle.\]
For $0\le t\le n$, and $y\in\{0, 1\}^t$, let
\[
q_y\defeq\sum_{z\in\{0, 1\}^{n-t}} p_{yz},\quad\textrm{and,}
\quad
 |\phi_t\rangle \ \defeq\  
\sum_{y\in\{0, 1\}^{t}} \sqrt{q_y} |y\rangle\otimes
|0\rangle^{\otimes n-t}.\]
Then $|\phi_0\rangle=|0\rangle^{\otimes n}$,
and $|\phi_n\rangle=|\phi\rangle$.

The algorithm has $n$ stages.
At the $\thnum{i}$ stage, the algorithm
transforms $|\phi_i\rangle$ to a state
$|\phi_{i+1}'\rangle$ such that
$\||\phi_{i+1}'\rangle-\phi_{i+1}\rangle\|\le\epsilon/n$,
and uses $\ell\defeq\lceil\log \frac{n}{\epsilon}\rceil$
basis-changing gates. This can be done by
the following.
\begin{enumerate}
\item 
For each $y\in\{0, 1\}^t$, let
$\theta_y\defeq \arccos(\sqrt{q_{y0}/q_y})$.
Compute on the ancilla and using Toffoli gates the first $\ell$
bits of $\theta_y/\pi$, $a_{y, 1}$, 
$a_{y, 2}$, $\cdots$, $a_{y, \ell}$.
This maps $|\phi_t\rangle\otimes|0\rangle^{\otimes \ell}$
to 
\[ \sum_{y\in\{0, 1\}^t} \sqrt{q_y} |y\rangle\otimes|0\rangle^{\otimes n-t}
\otimes|a_{y, 1}, a_{y, 2}, \cdots, a_{y, \ell}\rangle.\]
\item
Denote by $R(\theta)$ the single qubit rotation operator of
an angle $\theta$. Let $\theta_y'=\sum_{s=1..\ell} a_{y, s}\pi/2^s$.
For $s=1..\ell$, apply the Controlled-$R(a_{y, s}\pi/2^s)$ gate with
the $\thnum{s}$ qubit in the ancilla as the control qubit and the 
$\thnum{(t+1)}$
qubit in the output state as the destination qubit.
This results in mapping
\[
\sum_{y\in\{0, 1\}^t}\ \sqrt{q_y}\ |y\rangle\otimes|0\rangle
\ \to\ \sum_{y\in\{0, 1\}^t}\ \sqrt{q_y}\ |y\rangle\otimes R(\theta_y') |0\rangle.\]
Since $\|R(\theta_y')-R(\theta_y)\|\le \pi/2^\ell$,
and 
\[|\phi_{t+1}\rangle = \sum_{y\in\{0, 1\}^t} \sqrt{q_y}|y\rangle
\otimes R(\theta_y)|0\rangle\otimes|0\rangle^{\otimes n-t-1},\]
the resulted vector $|\phi_{t+1}'\rangle$ satisfies
\[\||\phi_{t+1}'\rangle-|\phi_{t+1}\rangle\| \le \pi/2^\ell.\]
\end{enumerate}

Hence, setting $\ell=\lceil\log_2 (\pi n/\epsilon)\rceil$,
the algorithm outputs a state $|\phi'\rangle\defeq|\phi'_{n+1}\rangle$ 
that satisfies
\[\||\phi'\rangle - |\phi\rangle\|\le
\sum_{t=1}^{n}\ \||\phi'_{t}\rangle - |\phi_t\rangle\| \le \epsilon.\]
The total number of basis-changing gates used is
$n\cdot\ell= O(n\log\frac{n}{\epsilon})$.
\end{proof}

We now 
consider the the general case.
\begin{theorem}
\label{thm:upper}
Let $|\phi\rangle$ be a quantum state
over $n$ qubits and $\epsilon>0$
be a constant. Then there exists a quantum
algorithm that uses 
$O(\frac{1}{\epsilon} H(\phi)\log \frac{H(\phi)}{\epsilon})
=O(H(\phi)\log H(\phi))$
number of basis-changing gates,
and maps $|0\rangle^{\otimes n}$
to a state $|\phi'\rangle$ such that
\begin{equation}\label{eqn:smalldist}
\||\phi'\rangle - |\phi\rangle\|\le \epsilon.
\end{equation}
\end{theorem}
\begin{proof}
Suppose for some $N>0$ and $\alpha_i\in[0, 2\pi)$, $p_i\ge0$, $0\le i\le N-1$,
\[ |\phi\rangle \ = \ \sum_{i=0}^{N-1}\ e^{i\alpha_i}\sqrt{p_i}\ |i\rangle,\]
where $\sum_{i=0}^{N-1} p_i=1$.

We first observe that the basis-preserving
gate 
\[G_\phi\defeq \sum_i e^{i\alpha_i}|i\rangle\langle i|\]
maps
\[ |\phi\rangle\ \to \ \sum_{i=0..N-1} \sqrt{p_i}\ |i\rangle,\]
and vice verser.
Therefore we can assume
that $\alpha_i=0$, for all $i$.

For a real $\lambda>1$ to be determined later,
define 
\[ W_\lambda \ \defeq\ \{ i : p_i\ge 2^{-\lambda H(\phi)}\},
\qquad p\defeq\sum_{i\not\in W_\lambda} p_i,\qquad\textrm{and,}\]
\[ |\phi_\lambda\rangle \ \defeq\ \sum_{i\in W_\lambda}
\sqrt{\frac{p_i}{1-p}} \ |i\rangle.\]
Then we have $p\le\frac{1}{\lambda}$. Hence,
$|W_\lambda|\le 2^{\lambda H(\phi)}$, and 
\[ \||\phi\rangle -|\phi_\lambda\rangle\|\le p\le\frac{1}{\lambda}.\]

Now set $k\defeq\lambda H(\phi)$.
After an appropriate permutation $\sigma$ on $\{0, 1\}^k$,
$|\phi_\lambda\rangle$ can be written as
\[ |\psi_\lambda\rangle\defeq \sigma|\phi_\lambda\rangle
=\sum_{x\in\{0, 1\}^k} \sqrt{q_x} |x\rangle.\]
By Lemma~\ref{lm:special}, we can generate a state
$|\psi'_\lambda\rangle$ using $O(k\log (\lambda k))$
basis-changing gates and 
\[\||\psi'_\lambda\rangle
-|\psi_\lambda\rangle\|\le \frac{1}{\lambda}.\]
The output state is 
$|\phi'_\lambda\rangle\defeq\sigma^{-1}|\psi'_\lambda\rangle$,
which satisfies
\[\||\phi'_\lambda\rangle-|\phi\rangle\|\le
\||\phi'_\lambda\rangle - |\phi_\lambda\rangle\|
+ \||\phi_\lambda\rangle - |\phi\rangle\|
\le \|\psi'_\lambda\rangle - |\psi_\lambda\rangle\|
+1/\lambda\le 2/\lambda.\]
Setting $\lambda=2/\epsilon$, this gives the required
precision. The total number of basis-changing gates
used is 
\[O(k\log(\lambda k)) =O(\frac{1}{\epsilon} H(\phi)\log \frac{H(\phi)}{\epsilon})
=O(H(\phi)\log H(\phi)).\]
\end{proof}

\begin{remark}[Improving the upper bound]
\label{rmk:close}
Since for some quantum states $|\phi\rangle$, a small perturbation may
reduce $H(\phi)$ dramatically, the upper bound
in Theorem~\ref{thm:upper} may be improved
by approximating such a lower entropy approximation state.
For example, consider
\[ |\phi_\delta\rangle = (1-\delta)|0\rangle
+\sum_{i\in[K]} \frac{\sqrt{2\delta-\delta^2}}{\sqrt{K}} |i\rangle.\]
Then $H_{\phi_\delta} = \Theta(\log K)$. On the other hand,
if $\epsilon\ge 2\delta$, the constant state $|0\rangle$
is an $\epsilon$-approximation of $|\phi_\delta\rangle$.
Hence no basis-changing gate is needed at all.
\end{remark}
\subsection{Lower bound.}
We can generalize the definition of $H(\phi)$
to $H(\rho)$ for a mixed state $\rho$ in the obvious way.
Denote the trace norm of a matrix $M$ by $\|M\|_{tr}$.
Given a state $|\phi\rangle$ and a real $\epsilon>0$
let 
\[ H_\epsilon(\phi)\ \defeq \ \inf \ 
\{ H(\rho) : \|\rho - |\phi\rangle\langle\phi|\|_{tr}\le \epsilon\}.\]
Note that $H_\epsilon(\phi)$ could be substantially smaller
than $H(\phi)$, as demonstrated by the example in 
Remark~\ref{rmk:close}. On the other hand,
for some family of states, such as the uniform superpositions
$\{ \frac{1}{\sqrt{N}}\sum_{i=0..N-1} |i\rangle : N>0\}$,
$H_\epsilon(\phi) = \Theta(H(\phi))$, for small $\epsilon$.

\begin{theorem}
\label{thm:lower}
Let $|\phi\rangle$ be a quantum state and $\epsilon>0$ be a constant.
Then any quantum algorithm that generates a mixed state $\rho$
that satisfies $\|\rho -|\phi\rangle\langle \phi|\|_{tr}\le \epsilon$ must use
$\Omega(H_\epsilon(\phi))$ number of basis-changing gates.
\end{theorem}

Notice that if $\||\phi\rangle-|\phi'\rangle\|\le\epsilon$,
then 
\[\||\phi\rangle\langle\phi|-|\phi'\rangle\langle\phi'|\|_{tr}
\le 2\epsilon.\]
Therefore, in general, the algorithm in Theorem~\ref{thm:upper}
is almost tight (up to a logarithmic factor) for sufficiently
small $\epsilon$ and family of states that have $H_\epsilon(\phi)
=\Theta(H(\phi))$.
\begin{proofof}{Theorem~\ref{thm:lower}}
Suppose $k$ number of basis-changing gates are used
to generate $\rho$. We will prove that $k=\Omega(H_\epsilon(\phi))$.

Denote the state after the $\thnum{i}$
basis-changing gate by $|\phi_i\rangle$, $0\le i\le k$.
Note that
$\rho=F(|\phi_k\rangle\langle \phi_k|)$, for some physically realizable operator
$F$, which is a composition of a permutation  (with some phase)
of the computational
basis followed a partial trace. Hence 
\begin{equation}
\label{eqn:last}
H(\phi_k)\ge H(\rho) \ge H_\epsilon(\phi).
\end{equation}
Since $H_0=0$, it suffices to prove that 
$H_{i+1} \le H_i+ C$, for 
all $0\le i\le k-1$, and some constant $C$.

Fix a $t$, $0\le t \le k-1$.
Let $U_t$ be the $\thnum{t}$ basis-changing gate,
which is applied to a set of qubits $A$.
The other qubits are denoted by $B$. 
Note that the number of qubits in $A$, denoted by $C$, is a constant.
Denote by $A'$ and $B'$ two new systems that have the same number
of qubits as in $A$ and $B$, respectively.
Let $Copy[A; A']$ be the product of Controlled-Not gates
that use qubits in $A$ as the control and the corresponding
qubits in $A'$ as the destination. Similarly define
$Copy[B; B]$. 
\newcommand{\rhot}[1]{|\psi_{#1}\rangle\langle\psi_{#1}|}
Let 
\[ |\psi_t\rangle\ \defeq\ 
 \left( Copy[A; A] \cdot Copy[B; B]\right)\  
|\phi_t\rangle_{AB}\otimes|00\cdots 0\rangle_{A'B'}.\]
Define $|\psi_{t+1}\rangle$ similarly.
Denote the von Neumann entropy of a mixed state by $E(\cdot)$.
Then
\[ H_t= E\left((\rhot{t})_{AB}\right)
= E\left((\rhot{t})_{A'B'}\right).\]
The second equality follows from that $|\psi_t\rangle$
is a pure state. Similarly,
\[H_{t+1}=E\left((\rhot{t+1})_{AB}\right)
= E\left((\rhot{t+1})_{A'B'}\right).\]

By the subadditivity of von Neumann entropy,
\[ E((\rhot{t+1})_{A'B'})\le
E((\rhot{t+1})_{A'})+E((\rhot{t+1}_{B'}).\]
Since 
\[|\psi_{t}\rangle\ =\ 
\left( Copy[A; A']\cdot U_t^\dagger\cdot Copy[A; A']\right)
\ |\psi_{t+1}\rangle,\]
we have
\[E((\rhot{t+1})_{B'})=E((\rhot{t})_{B'}).\]
The latter is exactly 
\[ E((\rhot{t})_{ABA'}\le E((\rhot{t})_{AB}) + E((\rhot{t})_{A'}),\]
by the subadditivity again.
Putting the above together, we have
\[ H_{k+1}\le H_k + E((\rhot{t})_A) + E((\rhot{t+1})_{A'})
\le H_k + 2C.\]
Together with (\ref{eqn:last}), this implies
$k=\Omega(H_\epsilon(\phi))$.
\end{proofof}
\section{Quantum and Classical Tradeoffs in solving Grover's Problem.}
In this section, we prove a quantum and classical
tradeoff relation for Grover's Problem \cite{Grover:1996:Search},
which is also called
Unstructured Search Problem. 
We start with the framework in which Grover's Problem is formulated
and then present the main result.
\subsection{Grover's Problem.}
The input to Grover's Problem (or, the Unstructured Search Problem) of size
$n$ is a binary string 
$x=x_0 x_2 \cdots x_{n-1}$,  where $x_i\in\{0, 1\}$, $0\le i\le n-1$, with 
the promise that there exists one and only one index $i$ such
that $x_i=1$. The task is to identify $i$. The complicacy is that
$x$ is known {\em only} to an oracle, which can only be accessed by
applying the oracle gate $O_x$:
\[ O_x |i, b\rangle = |i, b\oplus x_i\rangle,\quad 0\le i\le n-1,\ 
b\in\{0, 1\}.\]

Hence, in general, an algorithm would start
with a constant vector $|\phi_0\rangle$ in its state space, 
apply a sequence of unitary transformations
$U_0$, $O_x$, $U_1$, $O_x$, $\cdots$, $O_x$, $U_T$,
which is followed by a measurement
that would output $i$ with a high probability
(say $\ge 2/3$).
The complexity of the algorithm is $T$, the number of applications of
$O_x$.

In one of the most important papers in quantum
computing, Grover \cite{Grover:1996:Search} discovered a surprising quantum
algorithm that makes only $O(\sqrt{n})$ queries, a quadratic
speedup over the best possible classical algorithm. Because     
Grover's Problem is formulated in such a general way,                               
Grover's Algorithm can be used in solving many other problems
with a quantum speedup. A recent example is Ambainis' quantum 
algorithm for the classical problem of Element Distinctness
\cite{Ambainis:2003:ele}.                   
In fact, Grover's Problem is an example of problems formulated
in the so-called ``black-box model'', which has been widely
studied by many authors (see, e.g., the survey of Ambainis
\cite{Ambainis:2001:survey}).

\subsection{Quantum and classical tradeoffs for Grover's Problem.}
Much work has been done on proving lower bounds in the quantum
black-box model (see, e.g., two representative papers
by Beals, Buhrman, Cleve, Mosca, and  de Wolf \cite{Beals:2001:QLB},
and by Ambainis \cite{Ambainis00}).
In fact, the tight lower bound for Grover's Problem
was known before Grover's work due to Bennett, Bernstein,
Brassard, and Vazirani \cite{BennettBBV97},
and was refined by Boyer, Brassard, H{\o}yer, and Tapp
\cite{Boyer96}, and by Zalka \cite{Zalka99}.
\begin{theorem}[\cite{BennettBBV97}]
\label{thm:Groverlb}
Any quantum algorithm for solving Grover's Problem
of size $n$ must query $\Omega(\sqrt{n})$ times.
\end{theorem}

A quantum black-box algorithm can viewed as
a sequence of blocks of classical reversible computation that
may include oracle queries and are separated by
layers of basis-changing gates.
For example, for some $T=\Theta(\sqrt{n})$,
Grover's Algorithm uses $2T+O(1)$ Fourier transforms,
and in between,
$T$ oracle queries together with other classical
reversible computation.
We are interested in the tradeoff of the number of 
basis-changing layers and the number of queries.
\begin{theorem} \label{thm:tradeoff}
Any quantum algorithm solving Grover's problem
of size $n$
using $T$ queries and $\ell$ Fourier transforms must satisfy
$T\cdot\ell=\Omega(n)$.
\end{theorem}

A special case where the algorithm is required
to make $s$ queries non-adaptively, for a fixed $s$,
before making a local computation
was studied by Zalka \cite{Zalka99}, which
implies the same lower bound as the above for this case.

It is not hard to see that 
this tradeoff relation is optimal as long as $T=\Omega(\sqrt{n})$:
\begin{proposition}\label{prop:upper}
For any $T\ge \sqrt{n}$, there exists a quantum
algorithm that solves Grover's Problem of size $n$
 using $\Theta(T)$ queries and $\Theta(n/T)$ layers
of basis-changing gates.
\end{proposition}

\subsection{Proofs.}
We shall prove Theorem~\ref{thm:tradeoff} by a generalized form
of the ``quantum adversary''
technique of Ambainis \cite{Ambainis00}, which 
we now briefly review.

Let $f$ be a function defined on two disjoint sets 
$X$ and $Y$, where $X,\ Y\subseteq\{0, 1\}^n$, 
and for any pair $x\in X$, and $y\in Y$,
$f(x)\neq f(y)$. Let $R\in X\times Y$, and 
\[ m\ \defeq\  \min_{x\in X}\  \left| \{
y : (x, y)\in R\}\right|,\qquad
 \ell\ \defeq\ \max_{x\in X, i\in[n]}\ \left|\{
y : (x, y)\in R\ \textrm{and}\ x_i\neq y_i\}\right|,\]
and $m'$ and $\ell'$ are defined similarly 
with $X$ ($x$) and $Y$ ($y$) switched. Then 
\begin{lemma}[\cite{Ambainis00}]
\label{lm:ambaninis} Any quantum algorithm
that computes $f$ with error probability $\le\epsilon$,
$0\le \epsilon<1/2$,
must make $\Omega(\sqrt{mm'/\ell\ell'})$ queries.
\end{lemma}

This can be proved by considering the changes
on a ``progress indicator'' after each query of the algorithm.
Specifically, suppose we fix an algorithm that makes 
$T$ queries. Let $|\phi_z^t\rangle$ be the state
with oracle $z$ and after the $\thnum{t}$ oracle query.
Define the progress indicator
\[ p_t\ \defeq\ E_{x\in X, y\in Y} \ 
[\ \langle \phi_x^t | \phi_y^t \rangle\ ],\qquad t=0,\cdots, T.\]

Notice that only the oracle gate may change the progress indicator.
Clearly $p_0=1$. Furthermore, since the algorithm succeeds
with a probability at least $1-\epsilon>1/2$,
\begin{proposition}[\cite{Ambainis00}] 
For some constant $c$, $0\le c<1$, $p_T\le c$.
\end{proposition}
The lower bound is then established by
proving 
\[ |p_t - p_{t-1}|=O(\sqrt{\ell\ell'/mm'}),
\qquad\forall t\in[T]. \]

In our context, we shall consider the change
on the progress indicator $p_t$ after a sequence
of classical reversible computation with oracle queries.

\begin{lemma}\label{lm:generalize}
Let $f$, $X$, $Y$, and $p_t$ be as described above.
Let $k\in[n]$, and
\[\alpha_k \ \defeq\  \max_{x\in X, s\subseteq [n], |s|=k} \frac{ 
|\{ y : \textrm{$(x, y)\in R$, $y$ differs from x when 
restricted to $s$}\}|
}{
|\{ y : \textrm{$(x, y)\in R$}\}|}.\] 
Similarly define $\beta_k$ with $x$
switched with $y$ and $X$ switched with $Y$.
Then for any $t$, after a sequence of classical 
reversible computation that uses $k$ queries,
\[ |p_t - p_{t+k}|=O(\sqrt{\alpha_k\cdot \beta_k}).\]
\end{lemma}

\begin{proof}
Denote the computational basis by $\mathcal{C}$.
Denote the starting state (before the sequence of 
classical reversible computation) with oracle $z$ by 
\[ |\phi_z\rangle = \sum_{c\in\mathcal{C}}\ \gamma_{z, c} |c\rangle.\]
For an input $z\in\{0, 1\}^n$, denote by $\sigma_z$
the permutation on the computational basis specified by 
the algorithm. 
Then after the classical reversible computation, 
$|\phi_z\rangle\to \sigma_z |\phi_z\rangle$.
Hence the change of the progress
indicator 
\[ |p_t-p_{t+k}| = \left|\ E_{x, y}\ [\langle \phi_x |\phi_y\rangle]
- E_{x,y}\ [\langle\phi_x| \sigma_x^\dagger \sigma_y |\phi_y\rangle]\ 
\right|\]
is upper bounded by
\begin{eqnarray}
&&E[ |\langle \phi_x | \sigma_x^{\dagger} \sigma_y - I | \phi_y\rangle|]\\
&\le&
\frac{1}{|R|}\sum_{(x, y)\in R}\sum_{c, c'} |\gamma_{x, c}|\cdot
|\gamma_{y, c'}| \cdot |\langle c |\sigma_x^{\dagger}\sigma_y - I |c'\rangle|\\
&\le&
\frac{1}{|R|} \sum_{\stackrel{(x, y)\in R,\ c,\ c'}{
c'\neq c,\ \sigma_y(c') = \sigma_x(c)}} |\gamma_{x, c}|\cdot
|\gamma_{y, c'}|\label{eq:first}\\
&&\quad+\frac{1}{|R|} \sum_{\stackrel{(x, y)\in R,\ c}{
\sigma_x(c)\ne\sigma_y(c)}} |\gamma_{x, c}|\cdot|\gamma_{y, c}|\label{eq:second}.
\end{eqnarray}

Let us bound the second summation (\ref{eq:second}) first. After applying
Cauchy-Schwartz, we have the upper bound
\begin{equation}
\label{eq:common}
 \frac{1}{|R|}\sqrt{\sum_{x, c} \sum_{\stackrel{y: (x, y)\in R}{
\sigma_x(c)\neq \sigma_y(c)}}\ |\gamma_{x, c}|^2}\cdot
\sqrt{\sum_{y, c} \sum_{\stackrel{x: (x, y)\in R}{
    \sigma_x(c)\neq \sigma_y(c)}}\ |\gamma_{y, c}|^2}.
\end{equation}
A fixed combination of
$x$ and $c$ determines a set of $k$ coordinates being queried.
If $y$ is identical to $x$ in these coordinates then
$\sigma_y(c)= \sigma_x(c)$. Therefore, with
\[m_x=\max_{s\subseteq[n], |s|=k} \ \left|\{
y : \textrm{$(x, y)\in R$ and 
$y$ differs from $x$ when restricted to $s$}\}\right|,\]
and $m_y$ similarly defined, 
the above equation is further upper bounded by
\[ \frac{1}{|R|}\sqrt{\sum_x m_x \cdot\sum_y m_y}\le \sqrt{\alpha_k\beta_k},\]
since $\frac{\sum_x m_x}{|R|}\le \alpha_k$
and $\frac{\sum_y m_y}{|R|}\le \beta_k$.

Now we bound the first summation (\ref{eq:first}).
By Cauchy-Schwartz, it is upper-bounded by
\[\frac{1}{|R|}\sqrt{
\sum_{x, c} \sum_{\stackrel{c',\ y: (x, y)\in R}{
c'\ne c,\ \sigma_x(c)=\sigma_y(c')}} |\gamma_{x, c}|^2}\cdot
\sqrt{
\sum_{y, c'} \sum_{\stackrel{c,\ x: (x, y)\in R}{
c\ne c',\ \sigma_x(c)=\sigma_y(c')}}
|\gamma_{y, c'}|^2}.\]

The constraints on $y$ and $c'$ in the first summation are equivalent to
that $\sigma_x(c)\ne \sigma_y(c)$ and $c'=\sigma_y^\dagger\sigma_x(c)$,
therefore the above is upper bounded by Equation \ref{eq:common},
hence by $\sqrt{\alpha_k\beta_k}$ as well.
Therefore, the change on the progress indicator is at most
$2\sqrt{\alpha_k\beta_k}$.
\end{proof}
\begin{remark} A lower bound better than $2\sqrt{\alpha_k\beta_k}$
is $\frac{2}{|R|}\sqrt{(\sum_x m_x)\cdot (\sum_y m_y)}$,
where $m_x$ and $m_y$ are defined in the above proof. 
However, for our purpose of proving Theorem~\ref{thm:tradeoff},
both bounds are the same.
\end{remark}

We are now ready to prove Theorem~\ref{thm:tradeoff}
\begin{proofof}{Theorem~\ref{thm:tradeoff}}
For the purpose of proving lower bound, 
it suffices to consider the following decision version of Grover's problem:
determine whether or not the oracle is
$e_0=0^n$ or $e_i$, the $n$ bit 
binary string that has the single $1$ at the $i$-th position,
for some $i\in[n]$. 
Let $f$ in Lemma~\ref{lm:generalize} be this decision
problem and set 
\[X\ \defeq\ \{\ e_0\ \},\qquad
Y\ \defeq\ \{\ e_i : 1\le i\le n\ \},
\qquad\textrm{and,}\qquad R\ \defeq\ X\times Y.\]

Fix an algorithm that makes $T$ queries and $\ell$
levels of basis-changing gates. Then the algorithm
can be divided into $\ell+1$ blocks of classical reversible
computation 
with the $\ell$ basis-changing layers separating them.
Number the blocks by $1, 2, \cdots, \ell+1$.
For each block $s$,
let $k_s$ be the number of queries in this block,
and $p_{s-1}$ be the progress indicator at the beginning
of the block. The progress indicator at the end of the
last block is denoted by $p_{\ell+1}$.
We have $\sum_{s=1}^{\ell+1} k_s=T$,
$p_0 = 1$, and $p_{\ell+1}\le c$ for some constant
$c$ with $0\le c <1$.

Furthermore, for each $k_s$,
\[ \alpha_{k_s} = \frac{k_s}{n},\qquad\textrm{and,}
\qquad
\beta_{k_s}=1.\]
Then, by Lemma~\ref{lm:generalize},
\[ |p_{s}- p_{s-1}| = O(\sqrt{k_s/n}),\qquad \forall s\in[\ell+1].\]
Hence
\[ \sum_{s\in[\ell+1]} \sqrt{k_s/n} = \Omega(1).\]
By the Cauchy-Swartz Inequality, the left hand side
is upper-bounded by 
\[ \sqrt{(\ell+1)\cdot (\sum_{s=1..\ell+1} k_s)/n}
\ = \Theta(\sqrt{\ell\cdot T/n}).\]
Hence $\ell\cdot T=\Omega(n)$.
\end{proofof}

Proposition~\ref{prop:upper} can be proved by using a
 mixture of classical exhaustive algorithm and Grover's algorithm.
\begin{proofof}{Proposition~\ref{prop:upper}}
Consider the following algorithm.
Divide the $n$ bits binary string into
$h\defeq\lceil (n/t)^2\rceil$ blocks.
Apply Grover's algorithm to search for a block that
contains the $1$, and within each block, query all
the bits. The total number of queries is
$\Theta(\sqrt{h}\cdot \frac{n}{h}) = \Theta(T)$,
and the total number of layers of 
basis-changing gates is $\Theta(\sqrt{h}) = \Theta(n/T)$.
\end{proofof}

\section{Discussion.}
We initiate the study of what we called ``quantum and classical tradeoffs'',
which in essence is the relation of the number
of basis-changing gates in a quantum circuit 
with the computation power of the quantum circuit.
Specifically, we prove lower and upper bounds
on the number of basis-changing gates for generating
a given quantum state, and prove an optimal
tradeoff relation between the number of a layers
of basis-changing gates and the number of queries
for algorithms that solve Grover's Problem.
We shall conclude this paper by formulating a class
of open problems in this direction.

Since Toffoli and Hadamard are universal for quantum computing
(see, e.g., Shi \cite{Shi:2003:BTH}),
we can assume that any quantum circuit involves only these
two gates.
Notice that the composition of a set of Hadamard gates is just
a Fourier transform over a tensor product of $\mathbb{Z}_2$.

For each integer $k\ge0$, define the complexity class 
$FH_k$ ($FH$ meant to stand for ``Fourier Hierarchy) to be 
languages that
can be decided with a bounded error probability by a quantum circuit of
polynomial size and $\le k$ Fourier transforms.
Notice that if only uniform families of quantum circuits are considered,
$FH_0=P$, and $FH_1=BPP$.
When $k=2$, $FH_2$ starts to have nontrivial quantum
computation power. For example,
the oracle version of $FH_2$ includes Simon's problem,
and Factoring can be done in $FH_2$ via Kitaev's Phase Estimation Algorithm.

It appears a reasonable conjecture that
in general, the number of Fourier
transforms can not be reduced without substantial increase
of the circuit size. 
\begin{conjecture}
For any $k\ge 0$, $FH_k\subsetneq FH_{k+1}$.
\end{conjecture}
Since we do not know how to 
prove strong lower bounds in a general model,
one may have to consider first oracle versions of the problem,
that is, show an exponential separation
between $FH_k$ and $FH_{k+1}$ relative to an oracle for any $k$.
Simon's Problem provides an oracle separation for 
$FH_1$ and $FH_2$. The iterated version of it, as well as
the Recursive Fourier Sampling problem in \cite{BennettBBV97}
appear to be good candidates for an oracle separation for a general
$k$.

\section{Acknowledgment}
I would like to thank Mike Mosca for pointing out that Factoring
can be done with two Fourier transforms via Kitaev's Phase Estimation
Algorithm, and the related work of Zalka. I am also indebted to
Serap Savari for a useful discussion regarding Shannon entropy.

\end{document}